\documentstyle[prl,aps,epsfig,floats]{revtex}
\def\DESepsf(#1 width #2){\epsfxsize=#2 \epsfbox{#1}}
\begin{document}
\pagestyle{empty}                                      
\preprint{
\hbox to \hsize{
\hbox{
            }
\hfill $
\vtop{
 \hbox{ }}$
}
}
\draft
\vfill
\twocolumn[\hsize\textwidth\columnwidth\hsize\csname 
@twocolumnfalse\endcsname
\title{\boldmath{Impact of a Light Strange-Beauty Squark on
 $B_s$~Mixing and Direct Search}}
\vfill
\author{Abdesslam Arhrib,	
 Chun-Khiang Chua and Wei-Shu Hou }
\address{
\rm Department of Physics, National Taiwan University,
Taipei, Taiwan 10764, R.O.C.
}

\date{\today}
%
%
\vfill
\maketitle
\begin{abstract}
If one has Abelian flavor symmetry, 
$s_R$-$b_R$ mixing could be near maximal. 
This can drive a ``strange-beauty" squark ($\widetilde{sb}_{1}$)
to be rather light, but still evade the $b\to s\gamma$ constraint.
Low energy constraints
imply that all other superpartners are at TeV scale,
except for a possibly light neutralino, $\widetilde \chi_1^0$.  
Whether light or heavy, the $\widetilde{sb}_{1}$
can impact on the $B_s$ system: 
$\Delta m_{B_s}$ and indirect $CP$ phase,
even for $B_s \to \phi\gamma$.
Direct search is similar to usual $\tilde b \to b\widetilde \chi_1^0$,
but existing bounds are weakened by 
$\widetilde{sb}_{1} \to s\widetilde \chi_1^0$ possibility.
All these effects could be studied soon at the Tevatron.
\end{abstract}
\vskip2pc]

\pagestyle{plain}
The source of $CP$ violation within the Standard Model~(SM)
rests in the flavor sector, which is not well understood.
With three quark generations, we have 6 masses, 
3 mixing angles and a unique $CP$ phase in 
the Cabibbo-Kobayashi-Maskawa (CKM) mixing matrix $V$.
Together with leptons,
the majority of SM parameters in fact lies in the flavor sector.
However,
the left-handed nature of weak dynamics screens out
the mixings and $CP$ phases (no longer unique) in 
the right-handed quark sector.
The actual number of flavor parameters are 
much larger than meets the eye!

The observed quark masses and mixings do,
however, exhibit an intriguing hierarchical pattern
in powers of $\lambda \equiv \vert V_{us}\vert$,
hinting at a possible underlying symmetry~\cite{horizontal}.
If this ``horizontal" or flavor symmetry is Abelian, 
then $s_R$-$b_R$ mixing would 
be near maximal \cite{Nir,chua},
although still hidden from view.
It is interesting that, 
if supersymmetry (SUSY) is also realized,
$\tilde s_R$-$\tilde b_R$ squark mixing could then be near maximal.
This could generate observable effects in $b\to s$ transitions
even if squark masses are at TeV scale \cite{Nir,chua}.
Furthermore,
one of the squarks, the ``strange-beauty" squark $\widetilde{sb}_{1}$,
could be driven by this large flavor violation
to be considerably below the other squarks~\cite{chua}.
Whether we have a light $\widetilde{sb}_{1}$ squark or not,
it is of great interest since 
the current bound on $\Delta m_{B_s}$ \cite{bosc}
indicates that it could be larger than SM expectations.

In this Letter 
we point out that a light $\widetilde{sb}_{1}$ squark
and a light neutralino $\widetilde \chi_1^0$
are allowed by the $b\to s\gamma$ constraint.
We explore the implications of 
large $\tilde s_R$-$\tilde b_R$ mixing
on $B_s$-$\bar B_s$ mixing and its $CP$ phase $\Phi_{B_s}$.
In case of light $\widetilde{sb}_{1}$,
we briefly comment on direct search.
All these effects can be covered at the Tevatron Run II,
which has just started.
Mixing dependent CP violation in $B_s \to \phi\gamma$ decay
can also be studied in the future.
We stress that, besides the assumptions of 
Abelian flavor symmetry and SUSY,
the quark mixing and $CP$ phase we study
are on similar footing as the usual CKM matrix.

Horizontal models try to
explain the mass and mixing hierarchies by
powers of $\lambda \sim \langle S\rangle/M$, 
where $\langle S\rangle$ is the expectation of a scalar field $S$ 
and $M$ is a high scale. 
For Abelian symmetries, 
the commuting nature of horizontal charges in general gives
$
M_{ij} M_{ji} \sim M_{ii} M_{jj}
$
 ($i$, $j$ not summed),
where ``$\sim$" indicates approximate
rather than exact equality.
This allows one to determine,
e.g.
$M_d^{32}$ from our knowledge of
$M_d^{22} \sim m_s \sim \lambda^2 m_b$, 
$M_d^{23} \sim V_{23} m_b \sim \lambda^2 m_b$ 
and $M_d^{33} \sim m_b$.
Hence \cite{Nir}
\begin{equation}
{\hat M_d} =  {M_d\over m_b}  \sim
\left[
\matrix{\lambda^4 & [\lambda^3] & [\lambda^3] \cr
         [\lambda^3] &\lambda^2 &\lambda^2 \cr 
         [\lambda] &1 &1}
\right],
\label{Mq}
\end{equation}
and similarly for $M_u$;
the $[\cdots]$ terms would be 
set to zero as explained shortly.
Diagonalizing $M_d$ by a biunitary $D_{L}$ and $D_{R}$ transform,
$D_{R}^{23} \sim 1$ is clearly the largest mixing element,
but its effect is hidden within SM.

Taking SUSY as
commuting with the horizontal symmetry, 
the squark mass matrices are fixed by 
the common horizontal charge of the chiral supermultiplet.
We take the usual approach that squarks are 
almost degenerate with common scale $\widetilde m$.
From Eq. (1) one finds that
$(\widetilde M^2_d)_{LR} = (\widetilde M^2_d)_{RL}^\dagger
 \sim \widetilde{m} M_d$,
$(\widetilde M^2_d)_{LL} \sim  \widetilde m^2 V$, while
\begin{equation}
(\widetilde M^{2}_d)_{RR} \sim
 \widetilde m^2 \left[
		\matrix{1 & [\lambda] & [\lambda] \cr
	         [\lambda] &1 &1 \cr 
        	 [\lambda] &1 &1}
\right],
\label{MRR}    
\end{equation}
where $(\widetilde M^{2}_d)_{RR}^{23,32} \sim \widetilde m^2$
if $s_R$ and $b_R$ have the same horizontal charge(s),
hence comparable to 
$(\widetilde M^{2}_d)_{RR}^{22,33} \sim \widetilde m^2$.

%
%
We are interested in the impact of 
$(\widetilde M^{2}_d)_{RR}^{23,32}$.
It is known that 4 texture zeros are needed~\cite{chua}
to fully evade the $\Delta m_K$ and $\varepsilon_K$ constraints.
Hence, we choose horizontal charges such that 
the $[\cdots]$ terms in Eqs.~(1) and (2) are all set to zero,
which is achievable under a U$(1)\times$U$(1)$ 
or higher horizontal group.
With the $d$ quark thus decoupled,
one is safe from all known low energy constraints.
However, one needs 
$(\widetilde M^2_u)_{LL}^{12} \sim \lambda \widetilde m^2$
to account for $V_{us}$~\cite{Nir}.
It is intriguing that $\widetilde m$, $m_{\tilde g} \sim $ TeV
brings~\cite{chua} $\Delta m_D$ right into the ballpark of 
current \cite{xD} experimental sensitivities.
This sets the scale for $\widetilde m$ and $m_{\tilde g}$,
for if they were lighter,
$\Delta m_D$ would be too large.
Similarly, $\Delta m_K$ constrains
$\tilde u_L$, $\tilde c_L$ and $\widetilde \chi^\pm$ loops,
implying also \cite{chua} that squarks are at TeV scale, 
while the wino part of the chargino is heavier than 500 GeV.

With $d$-flavor decoupled, the $s$-$b$ part of
$\widetilde M^{2}_{RR}$ in Eq.~(2) appears ``democratic".
More explicitly, one has
\begin{equation}
\widetilde M^{2(sb)}_{RR}  = 
\left[
\begin{array}{ll}
\widetilde m_{22}^2 &   \widetilde m_{23}^2 e^{-i\sigma} \\
\widetilde m_{23}^2 e^{i\sigma}  &  \widetilde m_{33}^2
\end{array}  \right]
  = R \left[
	\begin{array}{cc}
	\widetilde m_{1}^2 & 0 \\
	0 & \widetilde m_{2}^2
	\end{array}  \right]
    R^\dagger,
%
\end{equation}
in quark mass basis,
where $\widetilde m_{ij}^2 \sim \widetilde m^2$
are all $> 0$, and
\begin{equation}
R = \left[
	\begin{array}{cc}
	 c_\theta &   s_\theta \\
	-s_\theta e^{i\sigma}  &  c_\theta e^{i\sigma} 
	\end{array}  \right].
\end{equation}
The phase in $R$ absorbs 
the $\sigma$ phase in $\widetilde M^{2(sb)}_{RR}$,
which is on similar footing
as $\phi_3 \equiv \arg V_{ub}^*$~\cite{phi3}.
By nature of $\widetilde m_{23}^2 \sim \widetilde m_{22,33}^2$,
in general we have one suppressed eigenvalue $\widetilde m_{1}^2$
due to level splitting,
where $\theta$ is a measure of the
relative weight of $\widetilde m_{23}^2$
vs $\widetilde m_{33}^2 - \widetilde m_{22}^2$.
Since our case corresponds to
$\widetilde m_{22}^2 \simeq \widetilde m_{33}^2
 \simeq \widetilde m_{23}^2 \simeq \widetilde m^2$
because of Eqs.~(1) and (2),
near maximal mixing is implied.
The eigenstates hence carry both $s$ and $b$ flavors
and are called the strange-beauty squarks $\widetilde{sb}_{1,2}$.
Without much loss of generality, we take 
$\widetilde m_{22}^2 = \widetilde m_{33}^2 = \widetilde m^2$
(so $\theta = \pi/4$ and 
$\widetilde m_{1}^2 + \widetilde m_{2}^2 = 2\widetilde m^2$) 
and consider the ratio 
$\widetilde m_{23}^2/\widetilde m^2 \equiv 1 - \delta \simeq 1$. 
The squark mass eigenvalues 
must be positive to preserve color symmetry,
hence $\delta > 0$ is required.
For small $\delta$,
we have $\widetilde m_{1}^2 \cong \delta \, \widetilde m^2$
and $\widetilde m_{2}^2 \cong (2-\delta)\widetilde m^2$.
Thus, with some tuning, 
$\widetilde{sb}_{1}$ can become quite light, 
i.e. $\widetilde m_1^2 \ll \widetilde m_2^2 \simeq 2\widetilde m^2$,
the driving force being the large 
$(\widetilde M^2_d)_{RR}^{23,32}/{\widetilde m^2} \sim  1$ 
in Eqs.~(2) and (3).
We note that, assuming $\widetilde m \sim 2$ TeV,
tuning $\delta$ to $\lambda^2$, $\lambda^3$, $\lambda^4$
give $\widetilde m_1 =$ 440, 206, 97 GeV;
for $\widetilde m \sim 1$ TeV,
$\delta = \lambda$, $\lambda^2$, $\lambda^3$
give $\widetilde m_1 =$ 470, 220, 103 GeV.
In the following,
we limit ourselves to $\widetilde m_1 \geq$ 100 GeV.

Besides concerns about tuning,
the pressing question is that a 
light $\widetilde{sb}_{1}$ driven by large strange-beauty mixing
seems particularly dangerous in face of the $b\to s\gamma$ constraint. 
As shown in \cite{chua}, 
heavy squark and gluino loops are suppressed 
by $1/G_F\widetilde m^2$ compared to SM contribution,
such that $b\to s\gamma$ rate is hardly affected.
It is interesting that, 
{\it even with $\widetilde{sb}_{1}$ as light as 100 GeV,
the $b\to s\gamma$ constraint is still rather accommodating}.

Since mass splittings are large,
the calculation of short distance coefficients is done
following \cite{fran}.
The expressions for Wilson coefficients 
together with their renormalisation group equations (RGE)
can be found in \cite{buras,chua1}. 
Our model gives large RR and RL mixings,
while LL and LR mixings are suppressed by $\lambda^2$.
In terms of the loop-induced effective $bs\gamma$ couplings
$m_b\,\bar s [C_{7}\,R+C_{7}^\prime\,L]
 \sigma_{\mu \nu }F^{\mu \nu }b$,
it is $C_{7}^\prime$ that receives larger contributions.
This in itself provides some protection,
since $C_{7}^\prime$ is not generated in SM
($C_7^{\rm SM} \simeq -0.31$), 
hence  our SUSY effects enter $b\to s\gamma$ rate only quadratically.

We find that, although RL mixing is 
suppressed by $m_b/\widetilde m$,
its effect dominates over the RR contribution for $\cos\sigma <0$.
Let us first show that $C^\prime_{7RR}$ is finite and
suppressed by $m_{\tilde g}^2$ in the $\widetilde m_1^2\to 0$ limit.
By direct computation, 
one finds that the $\widetilde {sb}$-$\tilde g$ loop contribution
to $m_b\, C^\prime_{7RR}$ is proportional to
\begin{equation}
  \int dk^2 
    {k^4 \, m_b \, \widetilde m^2_1\, c_{\theta} s_{\theta} e^{-i\sigma}
      \over (k^2 + m^2_{\tilde g}) (k^2 + \widetilde m^2_1)^4}     
  - (\widetilde m^2_1\to \widetilde m^2_2),
\end{equation}
where ``super-GIM" cancellation 
is ensured by Eq. (4),
and the $\widetilde {sb}_2$ term decouples for heavy $\widetilde m_2^2$. 
Since RR mixing is chiral conserving, a factor of $m_b$ is needed,
while $\widetilde m^2_1 \, c_{\theta} s_{\theta} e^{-i\sigma}$ 
is from $(\widetilde M^2_d)^{23}_{RR}$.
The integral is clearly finite
in the $\widetilde m^2_1\to 0$ limit.
Using formulas from \cite{chua1}, we find
$C^\prime_{7RR}(M_{\rm SUSY}) \simeq 
-0.1 \, c_\theta s_\theta e^{-i\sigma}(0.8 {\rm \ TeV}/m_{\tilde g})^2$
for maximal super-GIM breaking
 (small $\widetilde m^2_1$, large $\widetilde m^2_2$) case.
The shift in $b\to s\gamma$ rate is
$\lesssim 2\%$ for $\widetilde m > m_{\tilde g} > 0.8$ TeV.

The $m_b/\widetilde m$ suppression of RL contribution is 
compensated by a chiral enhancement factor \cite{Fujikawa}
$m_{\tilde g}/m_b$ since chirality flip is via $m_{\tilde g}$.
With $\tilde b_L \to \widetilde {sb}_{1R}$ mixing
and $\tilde b_L$ heavy, 	
the $s_{\theta} e^{-i\sigma}$ factor in Eq.~(5) is replaced by 
$
(R^\dagger\widetilde M^2_{RL})^{23}/{\widetilde m^2}
\sim \left(c_\theta - s_\theta e^{-i\sigma}\right)\, m_b/\widetilde m,
$
where $R$ is given in Eq. (4), and we take
$(\widetilde M^2_d)_{RL}^{23,33}   
 \sim m_b \widetilde m$ as real.	
The factor $k^4/(k^2 + \widetilde m^2_1)^4$ in Eq. (5) is replaced by
$k^2/(k^2 + \widetilde m^2_1)^3$ and the integral is still finite
for $\widetilde m^2_1 \to 0$.
We find
$C^\prime_{7RL}(M_{\rm SUSY}) \simeq 
0.12 \, c_\theta (c_\theta - s_\theta e^{-i\sigma})
(1.6 {\rm \ TeV}^2/\widetilde m \, m_{\tilde g})$
for small $\widetilde m^2_1$.
Taking $s_\theta \sim 1/\sqrt{2}$, 
$C^\prime_{7RL}$ is subdominant for $\sigma \sim 0$, 
but dominates over $C^\prime_{7RR}$ for $\sigma \sim \pi$.

\begin{figure}[t!]
\smallskip  
\centerline{
            {\epsfxsize2.125in \epsffile{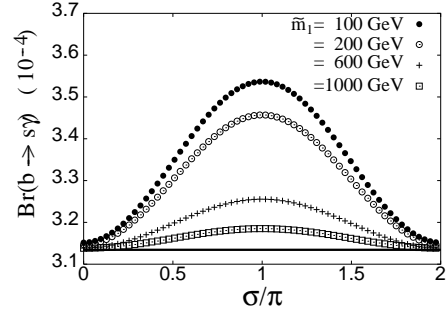}} }
\smallskip\smallskip\smallskip\smallskip\smallskip
\caption{$b\to s\gamma$ vs $CP$ phase $\sigma$  
including both SM and SUSY effects,
for $m_{\tilde g},\ \widetilde{m} = 0.8$, $2$ TeV 
and several strange-beauty squark mass
 ($\widetilde m_1 \equiv m_{\widetilde{sb}_{1}}$) values.
The horizontal line indicates
the SM expectation.}
 \label{fig:bsg}
\end{figure}

We illustrate in Fig.~1 the 
full gluino and neutralino loop effect on $b\to s \gamma$ rate
vs $CP$ phase $\sigma$,
for $m_{\tilde g}=0.8$ TeV and $\widetilde{m}=2$ TeV, 
with simplifying assumptions as stated above.
It is seen that, even for $\widetilde m_1$ as light as 100 GeV,
$b\to s\gamma$ is still \cite{limit} well within 
the allowed experimental range of 
$(3.15 \pm 0.54) \times 10^{-4}$~\cite{PDG}.
For heavy $\widetilde{sb}_1 \sim$ 1 TeV, 
its effect becomes negligible, and the $b \to s \gamma$ rate  
approaches the SM value, as 
indicated by the horizontal line at 
$b\to s\gamma$ $\sim 3.14 \times 10^{-4}$ for our parameter choice. 
The $\sigma$-dependence can be understood
through our earlier discussion of
$C^\prime_{7RR}$ and $C^\prime_{7RL}$.
One can also easily check 
from the strength of $\vert C_7^\prime\vert^2$ as seen in Fig. 1,
that the LR mixing contribution $\delta C_{7\rm LR}$ 
is indeed subdominant
even though it interferes with $C_7^{\rm SM}$ which is large.

It is intriguing that,
although $C_{7}^\prime$ is subdominant compared to $C_7^{\rm SM}$,
its strength is actually not small.
That is, $\vert C^\prime_7/C_7\vert \simeq 0.35 - 0.12$
hence $\sin2\vartheta = 2\vert C_7C_7^\prime\vert
                        /(\vert C_7\vert^2 + \vert C_7^\prime\vert^2)
                      \simeq 63\% - 22\%$ 
for $\widetilde m_1 = 100 - 1000$ GeV.
New physics effects \cite{chua,chua1} such as
mixing dependent $CP$ violation in $B^0\to K_1^0(1270)\gamma$
could be of this order 
(though direct $CP$ is small because $\delta C_{7\rm LR}$ is small),
but detectability may be better in $B_s \to \phi\gamma$. 
``Wrong" $\Lambda$-polarization in $\Lambda_b\to
\Lambda\gamma$ could also be promising \cite{chua1}.

It is known that charged Higgs effects on $b\to s\gamma$
add constructively to the SM  for all $\tan\beta$ \cite{HW},
giving rise to a very stringent constraint on $m_{H^+}$. 
Our light $\widetilde{sb}_1$ only worsens slightly the situation.
Taking 2$\sigma$ range of 
the measured $B \to X_s \gamma$ rate,
we find $m_{H^+} >$ 620, 660 (500, 600) GeV, respectively, 
for $\tan\beta=2$, $60$ and $m_{\tilde g} =$ 0.8 (1) TeV.
The heaviness of $H^+$ implies that 
{\it the second Higgs doublet is likely at the TeV scale as well}.

Turning to charginos, as stated, 
the $\Delta m_K$ constraint
demands that the wino part of chargino mass, 
controlled by $M_2$, should be larger than 500 GeV.
Because of stringent bounds from $b\to s\gamma$,
unless one makes fine-tuned cancellations \cite{Pokorski}
(e.g. with $H^+$ effect),
the higgsino part of chargino mass,
controlled by $\mu$, should also be at TeV scale,
especially for large $\tan\beta$.
We do not entertain a light stop
since we tacitly assume that 
flavor and SUSY scales are not too far apart \cite{Nir},
so the up squark mass average 
$\widetilde{m}_u$ is also at $\widetilde{m} \sim$ TeV. 
Thus, 
the charginos and 
the wino or higgsino-like neutralinos are all at TeV scale.
This still leaves open the possibility of a light
bino with mass controlled by $M_1$, which we call $\widetilde \chi_1^0$.
Interestingly,
$b\to s\gamma$ is not very constraining here:
we have taken the rather low mass value of  
$m_{\widetilde \chi_1^0} =$ 90 GeV in Fig.~1, 
and find that its effect is still much smaller than 
the dominant gluino contribution.  
This is simply because of the much weaker
bino coupling (hypercharge) to down sector
compared with the strong gluino couplings.

Without necessarily advocating a light bino,
we thus have a scenario
where SUSY particles and exotic Higgs bosons are at TeV scale,
except for a possibly {\it light neutralino $\widetilde \chi^0_1$
that is largely bino},
and a {\it light strange-beauty squark $\widetilde{sb}_1$
with mass driven low by flavor violation!}

One may worry that large $\tilde q$-$\widetilde{sb}_1$
(or $\widetilde \chi^-$-$\widetilde \chi_1^0$) splittings
may violate $\delta\rho$ constraint. 
We first note that $\delta\rho$ picks up
corrections to isovector gauge boson self-energy diagrams.
Our light bino case is hence of no consequence.
Because the isovector gauge interaction is left-handed, 
contributions from right handed squarks are
transmitted through LR mixing \cite{drees}. 
However, this is suppressed in our case 
by $\widetilde{M}_{LR}^2/\widetilde{m}^2
\sim m_{s,b}/\widetilde{m} \sim 10^{-4} - 10^{-3}$ \cite{edm}.
$\delta\rho$ can constrain only mass splittings in $\tilde{q}_L$,
which are TeV scale particles and do not have large splittings,
and thus the seemingly dangerous large splitting involving
$\widetilde{sb}_1$ is safe from $\delta\rho$ 
constraint.
We note in passing that our light $\widetilde{sb}_1$
can evade $R_b$ constraint also. 
The $\widetilde \chi^0$-$\tilde d_j$ contribution 
to $R_b$ is negligible \cite{Pokorski}
while $\widetilde \chi^-$-$\tilde t$ gives sizable contribution
only for light stop and light chargino,
which is not the case in our model.

Large $\tilde s_R$-$\tilde b_R$ mixing, however,
can easily impact on $B_s$-$\bar{B}_s$ mixing 
and its $CP$ phase $\Phi_{B_s}$, 
accessible soon at the Tevatron.
Recall that
$(\widetilde M^{2}_d)_{RR}^{23}/(\widetilde M^{2}_d)_{RR}^{13}
\sim 1/\lambda \sim \vert V_{ts}/V_{td}\vert$ in Eq. (2),
before setting [\ldots] terms to zero.
By simply scaling up the $B_d$ mixing results of \cite{chua} 
for $\tilde d_R$-$\tilde b_R$ mixing case,
one sees that even for $\widetilde m_1 \sim$ TeV,
its contribution to $B_s$ mixing could be of same order as SM.
The dominant $\tilde q$-$\tilde g$ box diagrams involve 
two $\widetilde{sb}_1$, 
or one $\widetilde{sb}_1$ and one $\tilde
s_L$/$\tilde b_L$ with $\tilde s_L$-$\tilde b_L$ mixing.
The former generates effective coupling 
$\propto \tilde C_1\, \bar s^\alpha_{R}\gamma_\mu b^\alpha_R\,
           \bar s^\beta_{R}\gamma^\mu b^\beta_R$,
while the latter
$\propto C_{4(5)} \, \bar s^\alpha_{R} b^{\alpha(\beta)}_L\,
           \bar s^\beta_{L} b^{\beta(\alpha)}_R$,
where $\tilde C_1 \propto c^2_\theta s^2_\theta e^{-2i\sigma}$, 
$C_{4(5)} \propto \lambda^2\, c_\theta s_\theta e^{-i\sigma}$ 
are known \cite{fran,tanaka} functions of 
$m_{\tilde g}^2/\widetilde m_1^2$ 
(simpler mass insertion formulas given in \cite{chua}).
Because of a larger loop factor,
the CKM suppressed $C_{4(5)}$ is comparable to $\tilde C_1$. 
Thus, the explicit $\sigma$-phase dependence
of the mixing amplitude is ($a$, $b$, $c$ are real)
\begin{equation}
M_{12} \equiv \vert M_{12} \vert e^{2i\Phi_{B_s}}
\cong a\, e^{-2i\sigma} + b\, e^{-i\sigma} + c,
\end{equation}
where $b$ (from $C_{4(5)}$) and $c$ (from SM) differ in sign.

\begin{figure}[t!]
\smallskip\smallskip    
\centerline{{\hskip0.352cm\epsfxsize1.5 in \epsffile{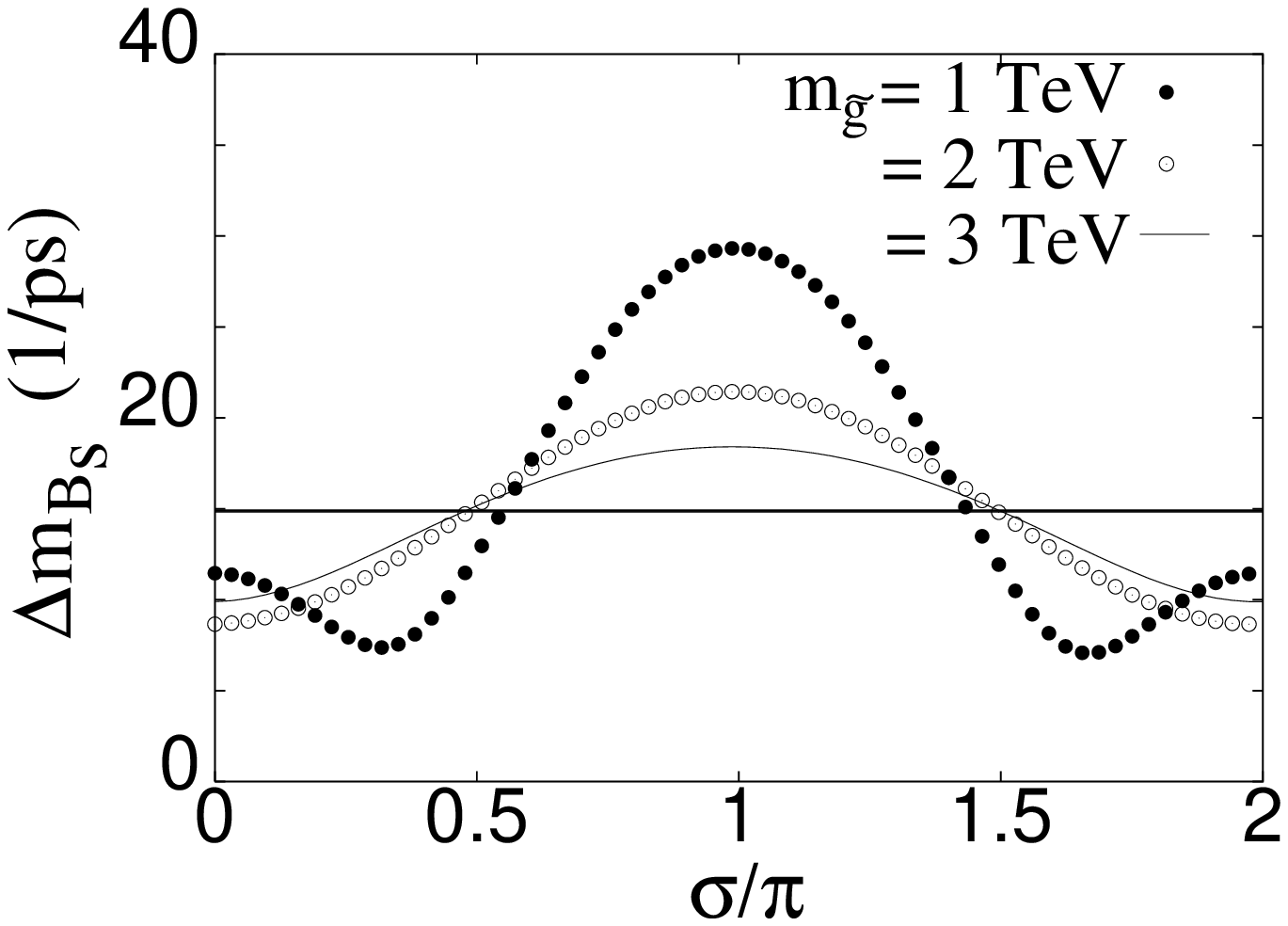}}
\hskip0.4cm
            {\epsfxsize1.5 in \epsffile{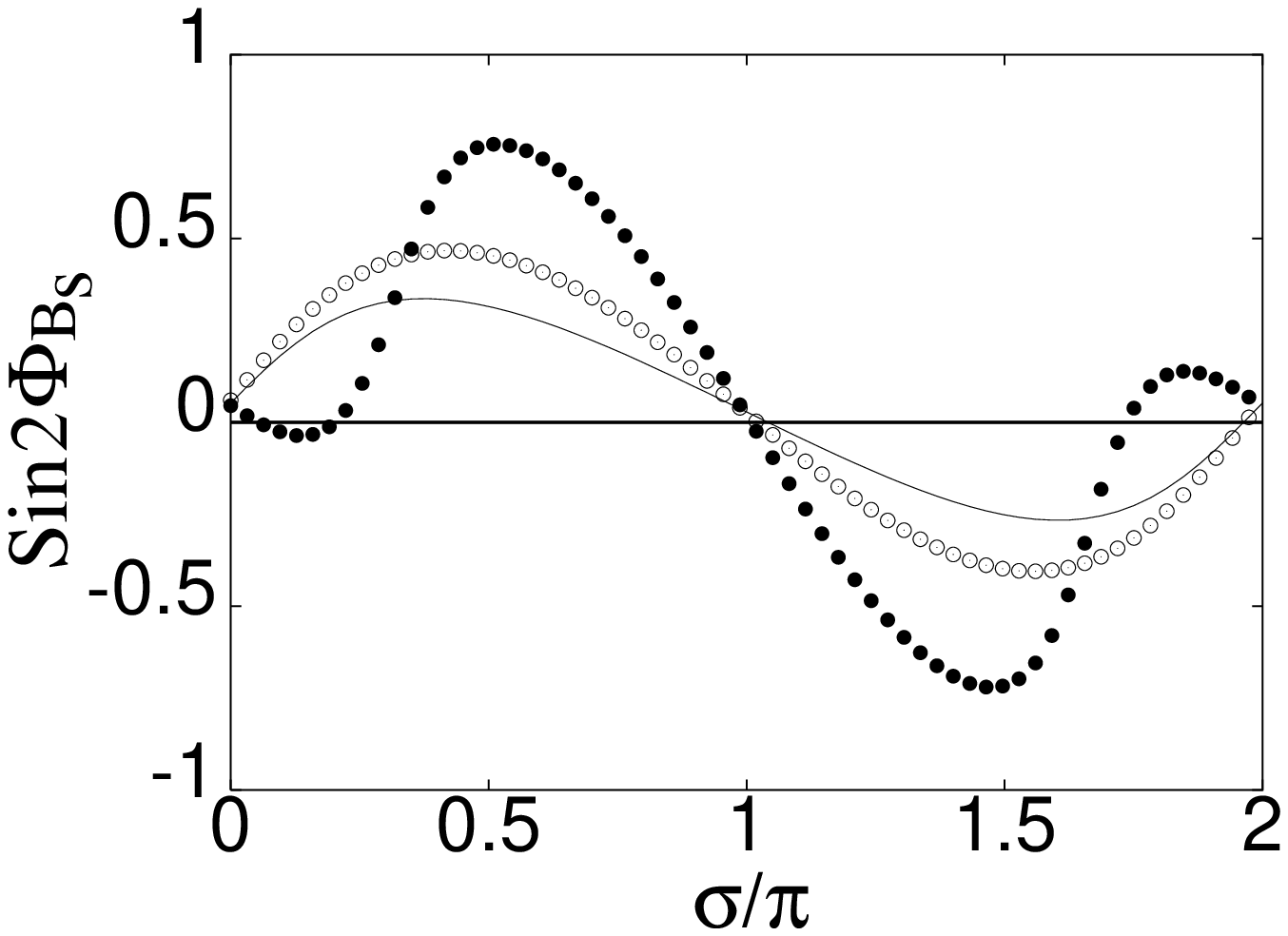}}}
\smallskip\smallskip\smallskip\smallskip
\caption{$\Delta m_{B_s}$ and  $\sin2\Phi_{B_s}$
vs $\sigma$ for $\widetilde m_{1}$, $\widetilde{m}=1.2$, 2 TeV
and $m_{\tilde g} =$1, 2, 3 TeV.
The horizontal line is SM expectation.
}
\label{fig:dbs2}
\end{figure}

Using RGE evolution from \cite{Bagger} and 
$f_{B_s}^2B_{B_s}=(240$ MeV$)^2$,
we find $\Delta m_{B_s}^{\rm SM} \simeq 14.9$ ps$^{-1}$
with vanishing $\sin2\Phi_{B_s}^{\rm SM}$.
For illustration, in Fig. 2 we plot 
$\Delta m_{B_s}$ and $\sin2\Phi_{B_s}$  vs $\sigma$ 
for $\widetilde m_1=1.2$ TeV, 
average squark mass $\widetilde{m}=2$ TeV
and $m_{\tilde g} =$ 1, 2 and 3 TeV.
As advertised, even for heavy $\widetilde{sb}_1$ at TeV scale,
the SUSY contribution can be comparable to the SM effect.
For $m_{\tilde g}=1$ TeV $< \widetilde m_1$, 
$\Delta m_{B_s}$ can reach twice the SM value around $\sigma \sim\pi$. 
For heavier $m_{\tilde g}$, 
$\Delta m_{B_s}$ can reach only  22 (18) ps$^{-1}$
for $m_{\tilde g}=$ 2 (3) TeV. 
Destructive interference between SM
and SUSY for $\cos\sigma > 0$ 
(where $\cos2\sigma$ modulation can be seen)
would give $\Delta m_{B_s} < \Delta m_{B_s}^{\rm SM}$
hence disfavored.
Thus, for the $\widetilde{sb}_1 \sim$ TeV scenario,
$\cos\sigma < 0$ is preferred.
Similarly, 
$\vert \sin2\Phi_{B_s}\vert$ can reach $50\%-75\%$,
vanishes at $\sigma = \pi$,
and has smaller range for heavier $m_{\tilde g}$.
If $\Delta m_{B_s}$ is only slightly above SM expectation,
it could be uncovered at
the Tevatron in a couple of years.
One could then find $\sin2\Phi_{B_s}\neq 0$
and indirect $CP$ in $B_s\to \phi\gamma$,
but no sign of SUSY particles since the scale is at TeV.

\begin{figure}[t!]
\smallskip
\centerline{
            {\hskip0.35cm\epsfxsize1.5 in \epsffile{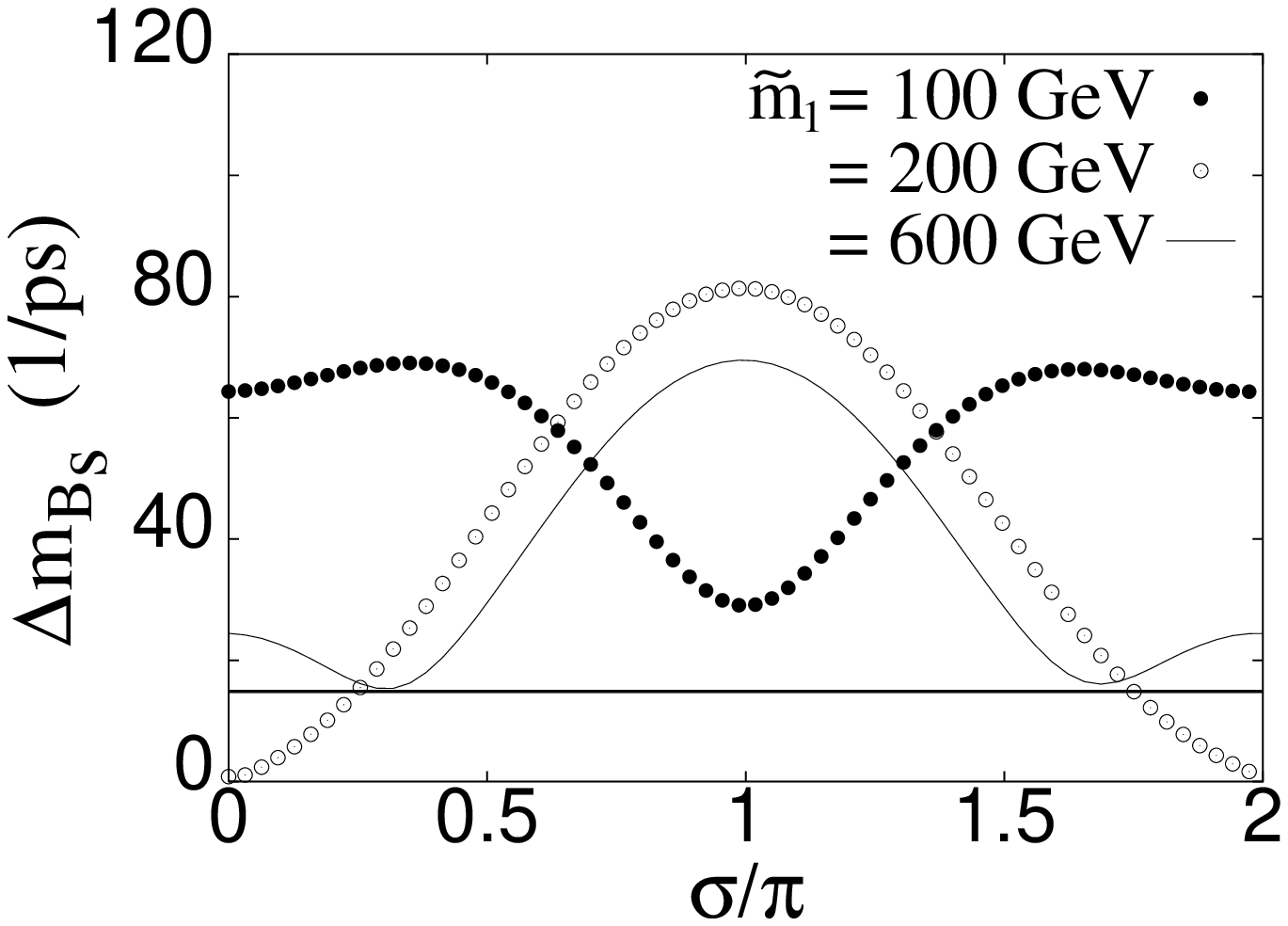}}
\hskip0.4cm
            {\epsfxsize1.5 in \epsffile{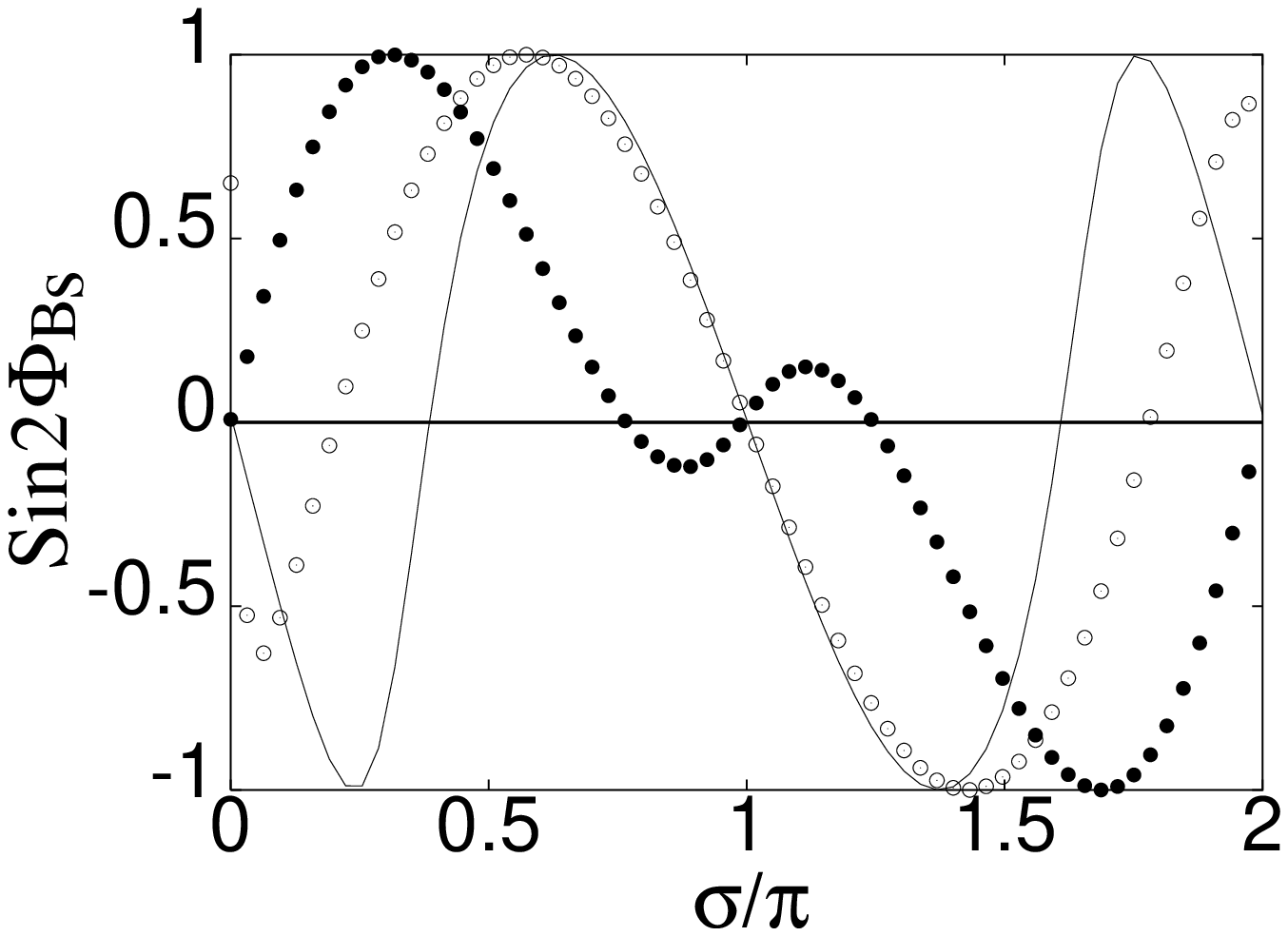}}}
\smallskip\smallskip\smallskip\smallskip
\caption {$\Delta m_{B_s}$ and $\sin\Phi_{B_s}$
vs $\sigma$ for $m_{\tilde g}$, $\widetilde{m}=0.8$, 2 TeV and
three $\widetilde m_1$ values.
The horizontal line is SM expectation.
}
\label{fig:dbs1}
\end{figure}

The light $\widetilde{sb}_1$ case allows greater range.
We plot $\Delta m_{B_s}$ and $\sin2\Phi_{B_s}$ vs $\sigma$ in Fig. 3,
for $m_{\tilde g}$, $\widetilde{m}=0.8$, 2 TeV,
and $\widetilde m_1 =$ 100, 200 and 600 GeV.
The $\widetilde{m}_1 = 600$ GeV case is similar to Fig. 2,
except that $a + b$ in Eq. (6) is of same sign as $c$.
For lower $\widetilde{m}_1$, 
the strength of $b$ increases monotonically
and is stronger than $c$,
while $a$ first drops slowly,
resulting in an accidental cancellation of $\Delta m_{B_s}$ 
at $\sigma = 0$ for $\widetilde{m}_1 \sim$ 200 GeV.
Below this, 
$a$ flips sign and changes rapidly,
and together with $b$ they overwhelm $c$.
Thus, for $\widetilde{m}_1 \lesssim 130$ GeV, 
one develops a dip rather than maximum at $\sigma \sim \pi$,
as shown for $\widetilde m_1 =$ 100 GeV case.

It is interesting that $\Delta m_{B_s}$ 
hovers not far above $15$~ps$^{-1}$
for both a broad range of $\widetilde m_1 \gtrsim 250$ GeV
and $\cos\sigma > 0$, 
and the intriguing case of
a rather light ($<$ 100 GeV) $\widetilde{sb}_1$
for phase $\sigma \sim \pi$.
For such $\Delta m_{B_s}$ values,
measurement would be swift,
with good prospects for $\sin2\Phi_{B_s}$,
which clearly covers the full range between $\pm 1$, 
with a $\sin2\sigma$ modulation over 
the basic $\sin\sigma$ dependence.
However, $\Delta m_{B_s}$ can also easily
reach beyond 40 ${\rm ps}^{-1}$,
whether $\widetilde{sb}_1$ is heavy or light,
and measurement would then take a while.
This in itself would indicate new physics,
but $\sin2\Phi_{B_s}$ measurement becomes difficult.
For confirming evidence, 
one would have to search for $C^\prime_7$ effects in $b\to s\gamma$, 
such as indirect $CP$ in $B_d \to K_1^0\gamma$ or
``wrong" $\Lambda$ polarization in $\Lambda_b \to \Lambda\gamma$.

Whether $\Delta m_{B_s}$ (and $\sin2\Phi_{B_s}$) is 
measured soon or not, it is imperative to check whether there is
a $\widetilde{sb}_1$ squark below a couple hundred GeV.
How should one search for it?
In the usual SUSY scenario, because of heaviness of top quark,
one could have a light stop
by RGE evolution from very high scale,
or by having large (flavor blind) LR mixing.
One could also have a light sbottom if  $\tan\beta$ is large.
This has motivated the experimental search \cite{tevlep}
via $\tilde{b}_1 \to b \widetilde{\chi}_1^0$
assuming that $\widetilde{\chi}_1^0$,
if not the lightest SUSY particle (LSP),
is lighter than $\tilde{b}_1$. 
The signature is two
b jets plus missing energy. 
In order to distinguish sbottom from stop,
$b$-tagging is necessary since loop-induced
$\tilde{t}_1 \to c \widetilde{\chi}_1^0$ leads to similar signature. 
In our case, all squarks including stop are at TeV scale,
except $\widetilde{sb}_1$
which becomes light because of large flavor violation,
{\it without the need for large $\tan\beta$}.
Since $\widetilde{sb}_1$ is a mixture of 
$\tilde s_R$ and $\tilde b_R$, 
both decays
$\widetilde{sb}_1 \to b \widetilde{\chi}_1^0$, 
$s \widetilde{\chi}_1^0$ are important,
and the $b$-tagging efficiency is diluted.
Thus, the standard sbottom search bound would weaken.
In any case, if a light sbottom is found, one would
have to check against production cross section
vs theory expectations from mass measurement,
to determine whether it is
the standard $\tilde b_1$ or the $\widetilde{sb}_1$.
In case $\widetilde{\chi}_1^0$ is heavier than $\widetilde{sb}_1$,
the LSP would likely be some sneutrino,
and the decay
$\widetilde{sb}_1 \to b \tilde\nu\nu$, $s \tilde\nu\nu$
via virtual $\widetilde \chi_1^0$ (hypercharge coupling)
has similar signature.


In conclusion,
flavor violation in $\tilde s_R$-$\tilde b_R$ squark sector 
could be uniquely large
if one has an underlying {\it Abelian flavor symmetry},
which are both inspired by the hierarchical patterns
of quark masses and mixings.
With SUSY above TeV scale,
this large flavor violation could evade low energy constraints,
including $b\to s\gamma$,
but modify $B_s$ mixing and generate $\sin2\Phi_{B_s} \neq 0$.
It is intriguing that
the strange-beauty squark $\widetilde{sb}_1$
{\it could be driven light by the large flavor violation itself}.
Both a light $\widetilde{sb}_1$ {\it and}
a light bino-like neutralino $\widetilde \chi_1^0$
can survive the $b\to s\gamma$ constraint.
This would not only further enrich $B_s$ physics,
but can also be directly probed via 
$\widetilde{sb}_1 \to b\widetilde \chi_1^0$, $s \widetilde \chi_1^0$,
which extends the standard
$\tilde b\to b\widetilde \chi_1^0$ search scenario.

This work is supported in part by
NSC-89-2112-M-002-063, 
NSC-89-2811-M-002-0086 and 0129, 
the MOE CosPA Project, 
and the BCP Topical Program of NCTS.


\begin{thebibliography}{99}


\bibitem{horizontal}
See, e.g. 
C.D. Froggatt and H.B. Nielsen, Nucl. Phys. {\bf B147}, 277 (1979).

\bibitem{Nir}
Y. Nir and N. Seiberg, Phys. Lett. {\bf B309}, 337 (1993);
M.~Leurer, Y. Nir and N. Seiberg, Nucl. Phys. {\bf B420}, 468 (1994).

\bibitem{chua} 
C.K. Chua and W.S. Hou, Phys. Rev. Lett. {\bf 86}, 2728 (2001);
and to be submitted.

\bibitem{bosc} Please see http://lepbosc.web.cern.ch/LEPBOSC/.

\bibitem{xD} R. Godang {\it et al.},
Phys. Rev. Lett. {\bf 84}, 5038 (2000);
J.M. Link {\it et al.},
Phys. Lett. {\bf B488}, 218 (2000). 


\bibitem{phi3}
The $\phi_3$ notation is preferred if 
impact of SUSY is found in $B$ system,
since $\beta$ and $\alpha$ are defined already.

\bibitem{fran} S. Bertolini {\it et al.}, 
Nucl. Phys. {\bf B353}, 591 (1991).

\bibitem{buras} A.J.~Buras {\it et al.},
Nucl.\ Phys.\ {\bf B424}, 374 (1994).

\bibitem{chua1} C.K. Chua, X.G. He, and W.S. Hou, 
Phys. Rev. D {\bf 60}, 014003 (1999).

\bibitem{Fujikawa}
K. Fujikawa and A. Yamada, Phys. Rev. D {\bf 49}, 5890 (1994);
P. Cho and M. Misiak, {\it ibid.} D {\bf 49}, 5894 (1994).

\bibitem{limit}
Even for $\widetilde m_1\to 0$,
$b\to s\gamma$ does not exceed $3.6\times 10^{-4}$.

\bibitem{PDG}  
D.E. Groom {\it et al.}, 
Eur. Phys. J. {\bf C15}, 1 (2000).

\bibitem{HW}
W.S. Hou and R.S. Willey, Phys. Lett. {\bf B202}, 591 (1988).

\bibitem{Pokorski} P.H. Chankowski and S. Pokorski, 
in {\it Perspectives on supersymmetry}, ed. G.L. Kane 
(World Scientific).

\bibitem{drees} M. Drees and K. Hagiwara, 
Phys. Rev. D {\bf 42}, 1709 (1990). 

\bibitem{edm} 
This allows us to evade the constraints of
D. Chang, W.Y. Keung and A. Pilaftsis,
Phys. Rev. Lett. {\bf 82}, 900 (1999).


\bibitem{tanaka} J.S. Hagelin, S. Kelley and T. Tanaka,
Nucl. Phys. {\bf B415}, 293 (1994).

\bibitem{Bagger} J.A. Bagger, K.T. Matchev and R.J. Zhang, 
                 Phys. Lett. {\bf B412}, 77 (1997).

\bibitem{tevlep} See e.g. T. Affolder {\it et al.}, 
Phys. Rev. Lett. {\bf 84}, 5704 (2000); 
%
G. Abbiendi {\it et al.}, 
Phys. Lett. {\bf B456}, 95 (1999).





\end{thebibliography}
\end{document}